\documentclass[
 preprint,
 amsmath,amssymb,
 aps,
]{revtex4-2}

\usepackage{graphicx}
\usepackage{dcolumn}
\usepackage{bm}

\begin{document}

\preprint{APS/123-QED}

\title{Momentum reconstruction from Unruh-deWitt detectors}

\author{Jesse Huhtala}
\email{jejohuh@utu.fi}
\author{Iiro Vilja}%
 \email{vilja@utu.fi}
\affiliation{%
 Department of Physics and Astronomy, University of Turku, Turku, 20014,  Finland
}%

\date{\today}

\begin{abstract}
We investigate momentum reconstruction for particle processes observed by Unruh-deWitt detector setups. In particular, we derive the probability distributions for particle momenta conditioned on detector clicks in three spatial dimensions. We investigate the statistical properties of such detector setups and discuss their use as models of measurement devices in particle physics.

\end{abstract}

\maketitle


\section{Introduction}\label{sec:intro}
Unruh-deWitt detectors have been used for a variety tasks: measurement theory for quantum field theory \cite{polo-gomezDetectorbasedMeasurementTheory2022,fraserNoteEpisodesHistory2023,martin-martinezWavepacketDetectionUnruhDeWitt2013}, relativistic quantum information and communication tasks \cite{anastopoulosQuantumFieldTheory2023,huRelativisticQuantumInformation2012,kinoshitaSpinSystemsQuantum2025,lapponiRelativisticQuantumCommunication2023,polo-gomezStateUpdatesUseful2025,simidzijaTransmissionQuantumInformation2020}, and entanglement harvesting \cite{chakrabortyEntanglementHarvestingQuantum2024, des.l.torresEntanglementStructureQuantum2023,maeso-garciaEntanglementHarvestingDetector2022,osawaEntanglementHarvestingQuantum2025,percheFullyRelativisticEntanglement2024}, among others. Given that a measurement theory for quantum field theories can be written in terms of Unruh-deWitt detectors \cite{polo-gomezDetectorbasedMeasurementTheory2022}, we think it would be interesting to replace outgoing scattering states with Unruh-deWitt detector states to investigate their suitability as particle detectors.

More specifically, our question is this: given that we know a particle process is happening, to what extent can we deduce the momenta of the particles thereby produced using Unruh-deWitt detector states? In particular, we will investigate the case of two outgoing massive scalar particles that were created as a result of the decay of a single massive scalar particle, which subsequently hit two Unruh-deWitt detectors. This is the old problem of momentum reconstruction, which is typically handled using multiple scattering theory \cite{betheMolieresTheoryMultiple1953, fruhwirthApplicationKalmanFiltering1987, glucksternUncertaintiesTrackMomentum1963a, highlandPracticalRemarksMultiple1975a, rossiCosmicRayTheory1941,scottTheorySmallAngleMultiple1963a}. However, we will not have to use machinery quite as heavy as that, since our detectors are mathematically simple and few in number.

Let us suppose that we choose interactions such that energy is conserved when a quantum moves from the field to the detector, that we know the incoming momentum of the decaying particle, and further that the detector is well-localized -- perhaps to a single point. Even with these helpful assumptions, reconstructing momenta is not a trivial task. The detectors will not conserve momentum, so how could we deduce momenta from them? 

The situation is not as hopeless as it seems. We first consider the problem of a classical particle of momentum $\mathbf{p}$ breaking apart in to two classical particles of smaller mass $m$ with momenta $\mathbf{k}_1$ and $\mathbf{k}_2$, which then hit point-like immobile detectors with excitation energies $\Delta_1$ and  $\Delta_2$, located at $\mathbf{x}_1$ and $\mathbf{x}_2$. The detectors absorb all the kinetic energy of the particles. We know the detectors have clicked. What can we deduce about the momentum of the classical particles in three spatial dimensions?

Momentum is conserved in the decay, so we know that $\mathbf{p} = \mathbf{k}_1+\mathbf{k}_2$. Further, we assumed that energy is conserved in the interactions, so either $E_{k_1} = \Delta_1$ or $E_{k_1} = \Delta _2$, and similarly for $E_{k_2}$. Further, a little geometric analysis shows that the two outgoing particles must be on the plane defined by the vectors $\{ \mathbf{p}, \mathbf{x}_1-\mathbf{x}_2\}$. We can make the problem two-dimensional by rotating the coordinate axes, and thus write the conserved quantities in component form as:
\begin{align}
    p_x &= k_{1x}+k_{2x},\\
    p_y &= k_{1y}+k_{2y},\\
    E_{k_1} &= \Delta_1 = \frac{|k_1|^2}{2m},\\
    E_{k_2} &= \Delta_2 = \frac{|k_2|^2}{2m}.
\end{align}
Note that since the labelling of the particles is arbitrary, we can choose either particle for each detector. This set of equations can be explicitly solved; the momenta of the outgoing particles are, in the classical case, fully determined.

The classical solution depends on well-defined deterministic particle trajectories. In the quantum case, we cannot uniquely determine a decay plane, as quantum fields do not admit sharp particle trajectories. By contrast, semi-classical wave packet arguments are a possibility to which we will return in section \ref{sec:filtering}. For now, we will instead have to be content with a partial determination of at least some of the momentum components. Our aim in the remainder of the article is to calculate to what extent momentum reconstruction can be achieved from Unruh-deWitt detectors coupled to a scalar QFT.

The basic idea of our strategy will be as follows. In the case of three spatial dimensions, we will have 6 unknown variables (the momenta of the outgoing particles). We will also have 5 constraints -- via delta functions -- from conservation laws: two from the detector energy conservation, and three from momentum conservation. We will integrate over these variables, and reconstruct 5 of the unknowns by using standard Dirac delta function rules. For the remaining variable we will obtain a conditioned probability distribution.

In the case of two spatial dimensions, it is possible to completely determine $\mathbf{k}_1$ and $\mathbf{k}_2$ from the constraints, as there are 4 constraints and 4 unknowns; the two-dimensional case thus basically reduces to the classical case. We are interested in the more general, non-trivial case, in particular the case of three space dimensions, to which we proceed forthwith. 
\section{Momentum reconstruction}\label{sec:momentumrec}
\subsection{Definitions}\label{sec:def}
We will use the decay of a massive scalar particle as the particle process studied with the Unruh-deWitt detectors. The outgoing field will be called $\Psi$ with mass $m$ and the incoming field will be $\phi$ with mass $M$. We call the dispersion relation of the $\psi$-particles $E_k$ and for the $\phi$-particles $\omega _p$. We will use the convention $\widetilde{dk} = \frac{d^Dk}{(2\pi)^D2E_{k}}$ where $D$ is the number of spatial dimensions. The particle process is mediated by the interaction Lagrangian
\begin{align}
    \mathcal{L}_{int} = -\lambda \int d^{D+1}x \phi(x):\Psi(x):^2
\end{align}
where $::$ denotes normal ordering.

To detect two particles, we need two Unruh-deWitt detectors, which are simply two-level systems with an energy gap $\Delta _i$. The detectors will be located at $x_1$ and $x_2$. They will both be coupled to a common source, such that either detector may have either of the energies. We will initially assume point-like detectors, though we will discuss other cases in section \ref{sec:discussion}. Thus, the interaction Hamiltonians are
\begin{align}
    H_{\text{det},i} &= -\epsilon_i \int dt \mu^{(i)} (t) \Psi (x_i,t),\\
    \mu^{(i)} (t) &=  \sum _{j\in \{1,2\}}[e^{-i\Delta _jt}b^{(i)}_j + e^{i\Delta _jt}(b^{(i)})^\dagger_j],
\end{align}
where $b_j/b^\dagger _j$ are the two-level annihilation/creation operators for energies indexed by $j\in\{ 1,2\}$. Both detector sites have the same two energy levels available. The index $i$ denotes the particle detector location. In accordance with this choice, the "detector click"\ state will be to a combination that erases which-way information:
\begin{align}
    |\Pi\rangle = \frac{1}{\sqrt{2}}\bigg[ |\Delta_1\rangle _{x_1}|\Delta_2\rangle _{x_2} + |\Delta_2\rangle_{x_1}|\Delta_1\rangle_{x_2} \bigg],
\end{align}
where $|\Delta _i\rangle_{x_j}$ means "the detector at $x_j$ had energy $\Delta _i$". We will discuss this unusual choice of detector state -- two Unruh-deWitt detectors in a superposition of energy eigenstates -- in section \ref{sec:discussion}. To recapitulate, our system consists of the Hilbert space
\begin{align}
    \mathcal{H} = \mathcal{F}_{\phi}\otimes \mathcal{F}_\Psi \otimes \mathbb{C}^4\otimes \mathbb{C}^4
\end{align}
with $\mathcal{F}_{\phi,\Psi}$ the Fock spaces for the scalar fields and $\mathbb{C}^4$ consists of two Unruh-deWitt detectors, $\mathbb{C}^4 = \mathbb{C}^2\otimes \mathbb{C}^2$.

We will assume that though both the detectors are on for a long time, there is no backreaction to the particle process; i.e. we assume the processes of particle production and detection as independent. We note that since $p \sim \Gamma T$ where $\Gamma$ is a rate per unit time, we can compute conditional probabilities using rates instead of probabilities.

The particle decay process gives for the rate to obtain momenta $k_1,k_2$ given an incoming momentum $p$:
\begin{align}
    d\pi (k_1k_2|p) = (2\pi)^{D+1} \delta^{(D+1)}(p-k_1-k_2)|\mathcal{M}(p,k_1,k_2)|^2\frac{1}{(2\pi)^D 2\omega_p}\widetilde{d^Dk_1}\widetilde{d^Dk_2}
\end{align}
where $\mathcal{M}$ is the scattering amplitude for the particle process and any other interactions that happen before the particles reach the detectors. For example, we leave for now open the possibility that the momentum distribution could be filtered by some additional part of the detector system. Additionally we get for $\dot{p}$, the probability (per unit time) for two particles to be absorbed in to the mixed two-detector state
\begin{align}
    \dot{p}(\Pi | k_1k_2) &= \mathcal{N}K(\mathbf{x}_1,\mathbf{x}_2,\mathbf{k}_1,\mathbf{k}_2)\bigg( \delta(E_{k_1}-\Delta _1)\delta(E_{k_2}-\Delta_2)+\delta(E_{k_1}-\Delta _2)\delta(E_{k_2}-\Delta_1) \bigg)
\end{align}
where
\begin{align}
    K(\mathbf{x}_1,\mathbf{x}_2,\mathbf{k}_1,\mathbf{k}_2) &= 2+2\cos ((\mathbf{k}_1-\mathbf{k}_2)\cdot (\mathbf{x}_1-\mathbf{x}_2)),
\end{align}
with $\mathcal{N} = \frac{\epsilon_1^2\epsilon_2^2}{2}$. This is derived in Appendix \ref{app:interference}.

\subsection{Three spatial dimensions}\label{sec:3d}
In three dimensions, we are interested in computing the probability distribution
\begin{align}
    p(\psi|\Pi, p) =\frac{1}{Z} \int \dot{p}(\Pi |k_1k_2)\delta(\psi-k_{1x})d\pi(k_1k_2|p),
\end{align}
where $k_{1x}$ labels component we anticipate will be left underdetermined after the conservation laws are used. As we will use spherical coordinates later, we call this component $\psi$. $Z$ is the witness, depending on $p$ and $\Delta _i$ but not on $k_i$. We have
\begin{align}
    Z = \int \dot{p}(\Pi |k_1k_2)d\pi(k_1k_2|p),
\end{align}
which is the total probability that the detectors are in the excited state. We assume that $\Psi$-particles could only have been produced by the prescribed decay process.

Since the other components are obtained by using the delta functions, we will obtain more or less explicit formulae for them \textit{en passant} while calculating the conditional distribution for $\psi$. 

We can now compute Z:
\begin{align}
    Z &= \int \dot{p}(\Pi |k_1k_2)d\pi(k_1k_2|p) \\
    &= \int \dot{p}(\Pi|k_1k_2)(2\pi)^4 \delta^{(4)}(p-k_1-k_2)|\mathcal{M}(p,k_1,k_2)|^2 \widetilde{d^3k_1}\widetilde{d^3k_2}
\end{align}
This integral can be computed explicitly, and we do this in appendix \ref{app:distribution}. The coordinate system will be rotated such that $\mathbf{p} \parallel z$. Define
\begin{align}
    E_k=\sqrt{m^2+k^2},
\qquad
\mathbf{r}=\mathbf{x}_1-\mathbf{x}_2,
\qquad
r=|\mathbf{r}|,
\qquad
P =|\mathbf{p}|.
\end{align}
Further, define the on-shell magnitudes obtained from integrating over the delta functions:
\begin{align}
    \kappa_i=\sqrt{\Delta_i^2-m^2},
\end{align}
and the polar angle $\theta$ of $\mathbf{k}_1$ with respect to $\mathbf{p}$ as
\begin{align}
\cos\theta_{ij}
=\frac{P^2+\kappa_i^2-\kappa_j^2}{2P\kappa_i},
\qquad
\sin\theta_{ij}=\sqrt{1-\cos^2\theta_{ij}}.
\end{align}
Then 
\begin{align}
    (2\mathbf{k}_1-\mathbf{p})\cdot\mathbf{r}
= A_{ij}+B_{ij}\cos\psi ,\quad K_{ij}(\psi) = \cos (A_{ij}+B_{ij}\cos(\psi)),
\end{align}
where 
\begin{align}
    A_{ij} &= 2\kappa_i r\,\cos\theta_{ij}\cos\alpha - Pr\cos\alpha, \\
B_{ij} &= 2\kappa_i r\,\sin\theta_{ij}\sin\alpha .
\end{align}
and $\alpha$ is the angle between $\mathbf{r}$ and $\mathbf{p}$. With these definitions, we show in appendix \ref{app:distribution} that
\begin{align}
    Z = \frac{\delta (\Delta E)\mathcal{N}}{(2\pi)^5\omega _p4P}\int_0^{2\pi} (K_{12}(\psi) + K_{21}(\psi))|\mathcal{M}(\psi)|^2d\psi,
\end{align}
with $\Delta E = \omega _p - \Delta_1 - \Delta_2$. Hence 
\begin{align}
    p(\psi | \Pi, p) = \frac{|\mathcal{M}(\psi)|^2(K_{12}(\psi)+K_{21}(\psi))}{\int_0^{2\pi} (K_{12}(\psi) + K_{21}(\psi))|\mathcal{M}(\psi)|^2d\psi}.
\end{align}
Supposing  that $\mathcal{M}$ does not depend on $\psi$, for example if it consists only of the prescribed particle decay process, we have the analytical result:
\begin{align}
    p(\psi|\Pi,p) 
=
\frac{
\displaystyle
\sum_{(i,j)\in\{(1,2),(2,1)\}}
\Big[1+\cos\!\big(A_{ij}+B_{ij}\cos\psi\big)\Big]
\,\Theta(\Delta_i-m)\,\Theta(1-|\cos\theta_{ij}|)
}{
\displaystyle
2\pi
\sum_{(i,j)\in \{ (1,2),(2,1)\}}
\Big[1+\cos(A_{ij})J_0(B_{ij})\Big]
\,\Theta(\Delta_i-m)\,\Theta(1-|\cos\theta_{ij}|)
}. \label{eq:distributionexplicit}
\end{align}
Thus we cannot exactly determine the final unknown angular momentum component, but we do obtain some information about it; the rest of the momentum components are determined by the conservation laws.

In fact, this result can be generalized to detectors that are not pointlike. Suppose that the coupling to the detectors is spatially extended, i.e.
\begin{align}
    H_{\text{det},i} = -\epsilon_i \int dtd^3x \mu(t) \chi_{x_i}(\mathbf{x})\Psi (t,x).
\end{align}
Then for centrosymmetric (about $x_i$) and real $\chi$, we can write
\begin{align}
    \int d^3x e^{i\mathbf{k}\cdot \mathbf{r}}\chi_{x_i}(\mathbf{x}) =e^{i\mathbf{k}\cdot \mathbf{x}_i}\tilde{\chi}_{x=0}(\mathbf{k}) := e^{i\mathbf{k}\cdot \mathbf{x}_i}\tilde{\chi}(\mathbf{k}).
\end{align}
Then we readily find
\begin{align}
    \dot{p}(\Pi | k_1k_2) &= \mathcal{N}|\tilde{\chi}(\mathbf{k}_1)|^2|\tilde{\chi}(\mathbf{k}_2)|^2K(\mathbf{x}_1,\mathbf{x}_2,\mathbf{k}_1,\mathbf{k}_2)\nonumber\\
    &\times \bigg( \delta(E_{k_1}-\Delta _1)\delta(E_{k_2}-\Delta_2)+\delta(E_{k_1}-\Delta _2)\delta(E_{k_2}-\Delta_1) \bigg).
\end{align}
As an example, for a spherical detector of radius $a$, we would have
\begin{align}
    \tilde{\chi}(\mathbf{k}) = \frac{4\pi a^3}{3}\frac{3j_1(|\mathbf{k}|a)}{\mathbf{k}a}
\end{align}
with $j_1$ the first spherical Bessel function. Let us define $\mathcal{F}_i = |\tilde{\chi}(\mathbf{k}_i)|^2$. Then we can see that
\begin{align}
    p(\psi | \Pi, p) = \frac{|\mathcal{M}(\psi)|^2(K_{12}(\psi)+K_{21}(\psi)))\mathcal{F}_1(\psi)\mathcal{F}_2(\psi)}{\int_0^{2\pi} (K_{12}(\psi) + K_{21}(\psi))\mathcal{F}_1(\psi)\mathcal{F}_2(\psi)|\mathcal{M}(\psi)|^2d\psi}
\end{align}
where the appropriate delta functions have been applied to $\mathcal{F}_i$.
\section{Statistical properties of the detector system}
Let us now investigate how well this detector system performs for determining $\psi$, the unknown momentum component, in the case of a point-like detector. Had we not used a superposition of energy eigenstates for the pair of Unruh-deWitt detectors, we would have obtained a uniform distribution on $\psi \in [0,2\pi]$. The uniform distribution also reveals the least possible information about $\psi$, so we will use it as a baseline. The motivation for using point-like detector is that such detectors provide a no-signaling measurement theory for quantum field theory, as shown by Polo-Gomez et al. \cite{polo-gomezDetectorbasedMeasurementTheory2022}.

We will first consider, in subsection \ref{sec:baredetector}, the absorptive detector setup defined by \eqref{eq:distributionexplicit}. We call this the "bare"\ detector. In the setup of the problem, we allowed that there might be a filtering process that further limits the momenta of particles arriving at the detectors, so that $\mathcal{M}(p\mapsto k_1k_2)$ consists of more than just the particle decay process. We will consider this possibility in subsection \ref{sec:filtering}. 

\subsection{Bare detector}\label{sec:baredetector}
It may be observed that the $\psi$-distribution \eqref{eq:distributionexplicit} is symmetric about $\pi$, but we will retain the full range of $\psi$ anyway. Further, the distribution must satisfy constraints: we must have $m < \Delta _i$ and
\begin{align}
    |\cos \theta_{ij}| = \bigg|\frac{P^2+\kappa_i^2-\kappa_j^2}{2P\kappa_i}\bigg| \leq 1.
\end{align}
In addition,
\begin{align}
    P^2 + M^2 = (\Delta_1 + \Delta_2)^2 \label{eq:total_energy_conservation}
\end{align}
and $M \geq 2m$, or else the decay process is not possible.

Let us begin our analysis from these constraints. When are they satisfied? The first one is easy to satisfy, since we can simply choose $\Delta _i$ such that they're larger than $m$. The cosine constraint is less obvious. After some algebra, one obtains
\begin{align}
    |\kappa _i - \kappa _j| \leq P \leq |\kappa_i + \kappa_j|.
\end{align}
This constraint is symmetric under $i\leftrightarrow j$, so either both products of Heaviside functions in \eqref{eq:distributionexplicit} equal one or neither. 

Once this constraint is satisfied, the energy conservation \eqref{eq:total_energy_conservation} can be satisfied for some appropriate $M\geq 2m$, simply by choosing $M$ so that the constraint is satisfied. One merely needs to make sure that $\Delta _1$ and $\Delta _2$ are not too small, i.e. that $\kappa _1$ and $\kappa _2$ are big enough. Our strategy is to simply choose some set of values $\kappa_1,\kappa_2,P,m$ and $M$ such that these constraints are satisfied, and investigate the properties of the detector as a function of the only available parameters aside from the energies -- their distance from one another and the angle of $\mathbf{r}$ with $\mathbf{p}$. Henceforth we assume that such parameters have been chosen.

First, we will analyze the case when $r \rightarrow \infty$, i.e. the large separation case. The distribution is of the form
\begin{align}
    p_r(\psi) = \frac{4 + 2\cos(r(a_{12}+b_{12}\cos (\psi))+2\cos(r(a_{21}+b_{21}\cos (\psi))}{N(r)} := \frac{p'_r(\psi)}{N(r)}
\end{align}
 where we have extracted a scaling factor $r$, with $a_{ij}=A_{ij}/r$, $b_{ij} = B_{ij}/r$, and $N(r) = \int_0^{2\pi} p'_r(\psi)d\psi$. Then our claim is as follows: the distribution $p(\psi)$ tends to the uniform distribution in the weak sense that
\begin{align}
    \int _0^{2\pi}f(\psi)p_r(\psi)d\psi \overset{r\rightarrow \infty}{\rightarrow} \frac{1}{2\pi}\int _{0}^{2\pi} f(\psi) d\psi
\end{align}
for a smooth test function $f(\psi)$. We see this as follows. Write $a_{ij}+b_{ij}\cos (\psi) = \Phi _{ij}$. Then
\begin{align}
    \int _0^{2\pi}f(\psi) p_r(\psi) d\psi = \frac{4 \int_0^{2\pi}f(\psi)d\psi + 2I_{12} + 2I_{21}}{N(r)} \label{eq:weakconv}
\end{align}
with $I_{ij} = \int_0^{2\pi} f(\psi) \cos(r \Phi_{ij})d\psi$.  Thus our task is to prove that $2I_{12}+2I_{21}$ approaches zero and $N(r) \rightarrow 1/8\pi$. However, as long as $b_{ij}\neq 0$, we have by the Lebesgue-Riemann lemma that $I_{12}$ and $I_{21}$ both go to 0 as $r\rightarrow \infty$. This immediately proves the claim. Note that the opposite limit, $r\rightarrow 0$, though degenerate, also results in a uniform distribution.

It is also possible to obtain by stationary-phase arguments that
\begin{align}
     \int _0^{2\pi}f(\psi) p_r(\psi) d\psi = \frac{4\int_0^{2\pi}f(\psi)d\psi + \mathcal{O}(r^{-1/2})}{8\pi + \mathcal{O}(r^{-1/2})},
\end{align}
i.e. the distribution approaches the uniform one as $r^{-1/2}$.

For the finite-$r$ region, such estimates are not available. We will instead consider two statistical measures: the Shannon differential entropy
\begin{align}
    h(r) = -\int _0^{2\pi}p_r(\psi) \ln p_r(\psi)d\psi
\end{align}
and a "best guess"\ probability
\begin{align}
    M_\epsilon (r) := \max _{\psi _0}\int _{\psi_0-\epsilon}^{\psi_0+\epsilon} p_r(\psi)d\psi.
\end{align}
Note that as $r\rightarrow \infty$, $M_\epsilon(r) \rightarrow 2\epsilon/2\pi$, which is the result for a uniform distribution. $M_\epsilon (r)$ is the probability of finding a randomly drawn sample from $p(\psi)$ in the interval $[\psi_0 -\epsilon, \psi_0 + \epsilon]$, for whichever $\psi_0$ maximizes $M_\epsilon (r)$ for the given $r$.

\begin{figure}
    \centering
    \includegraphics[width=1.0\linewidth]{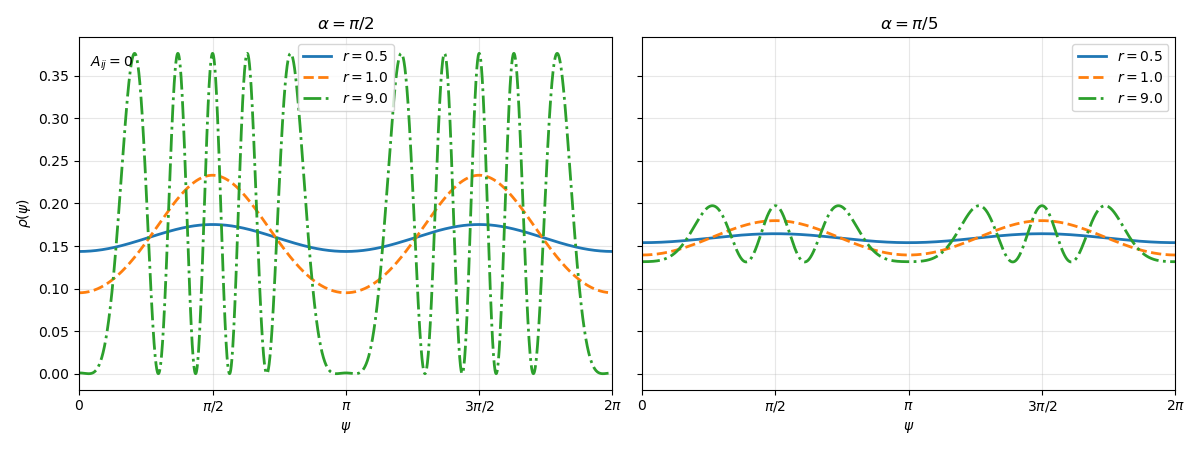}
    \caption{The probability distribution $\rho(\psi) := p_r(\psi)$ for an allowed set of particle parameters, and two given $\alpha$. Recall that $\alpha$ is the angle between $\mathbf{p}$ and $\mathbf{r}$. Note that if $\mathbf{p}$ and $\mathbf{r}$ are not perpendicular, the distribution is much close to uniform, as the incoming particle is exposed to a smaller cross-section of the detector setup. In the most degenerate case, $\alpha = 0$, the distribution no longer depends on $\psi$ at all.}
    \label{fig:psi_distribution}
\end{figure}

\begin{figure}
    \centering
    \includegraphics[width=1.0\linewidth]{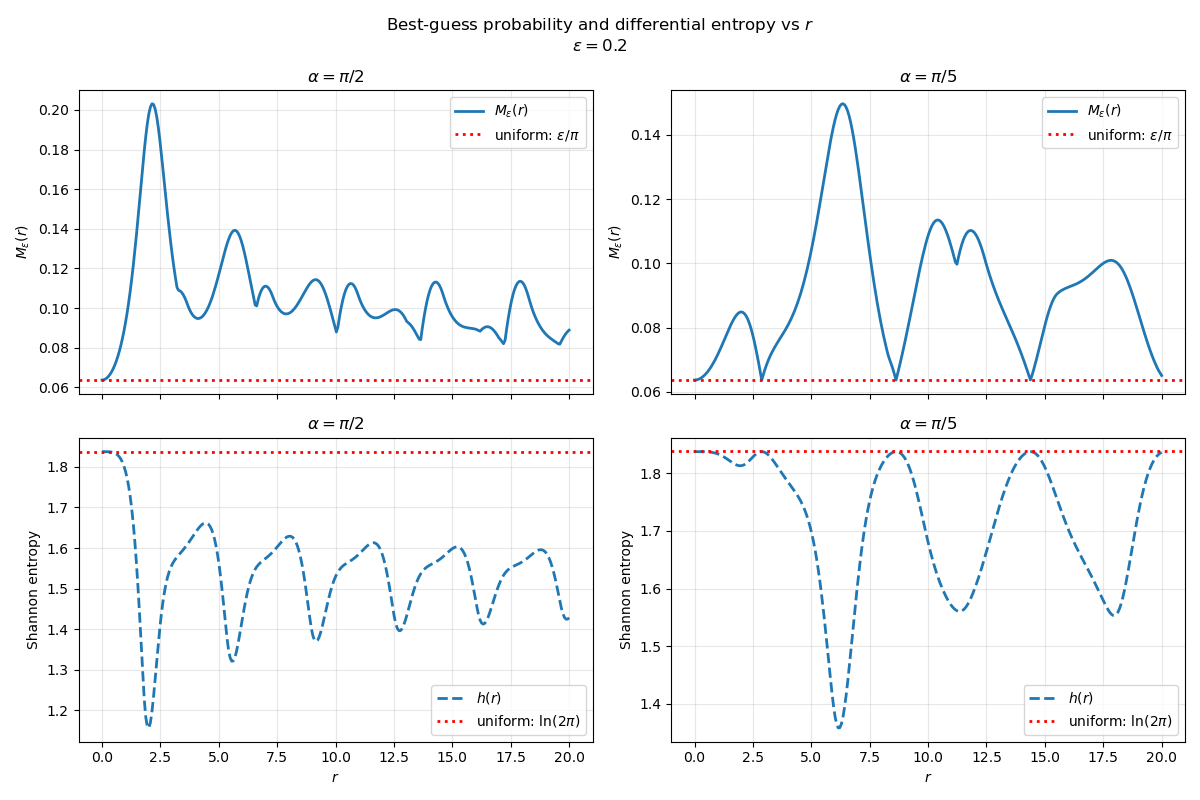}
    \caption{The Shannon differential entropy and best guess probability as functions of $r$. The dotted red line is the uniform distribution, which acts as a baseline. The optimal $r$ is clearly different for the differing values of $\alpha$, which can be understood intuitively: if $\mathbf{p}$ and $\mathbf{r}$ are not perpendicular, the cross section of the detector exposed to the approaching particles is smaller unless $r$ is increased.}
    \label{fig:statistical_markers}
\end{figure}

Given these definitions, we can choose some set of values $P, \kappa _1, \kappa_2, m$ and $M$ which satisfy the constraints and investigate the statistical properties as defined above. In figure \ref{fig:psi_distribution}, we have plotted the distribution; there appears to be, for each choice of $\alpha$, an optimal $r$ that provides the most information. This is shown more concretely in figure \ref{fig:statistical_markers} for our chosen statistical quantities.


What can we learn about the statistical properties of the bare detector system? Note that the magnitude of $B_{ij}$ is primarly determined by $r\kappa_i$. Realistic particle detectors have sensors meters apart. If we take a particle with an energy of 1 TeV, then the distance corresponding to $1/|\mathbf{k}|$ would be $O(10^{-19})$ meters in SI units. Thus, for a realistic particle detector, if we wanted $B \sim 10$, we would need the detector points to be $10^{-18}$ meters apart. Unruh-deWitt detectors therefore work best at separations many orders of magnitude smaller than what is present in realistic particle detectors. For realistic separations, we would essentially revert to a uniform distribution. Thus, while we can actually obtain a fairly good estimate of $\psi$ for optimal values of $r$, the detector parameters do not correspond to realistic detectors.
\subsection{Momentum filtering}\label{sec:filtering}
Our detector system was designed to consist of two separate steps: a particle production step (a decay process, in our case), and an absorption step, where the particles are absorbed and their energy is measured. We left open the possibility that the particle production step could contain also a filtering mechanism.

Since we are interested in the properties of Unruh-deWitt detectors, we will now illustrate what kind of filtering process could result from Unruh-deWitt detectors alone. Let us consider one of the earliest detector systems, a cloud chamber. There is a classic analysis of particle trajectories in a cloud chamber by Mott \cite{mottWaveMechanicsRay1929}. In the classic construction, one imagines the $\alpha$-particles entering the bubble chamber as interacting weakly with the hydrogen gas present. A perturbative analysis is applied to second order to find the excitation probabilities for the hydrogen, and Mott finds that only those parts of the wave function parallel to the position vector between two hydrogen atoms survive.

A similar analysis can be applied using Unruh-deWitt detectors. Take two ordinary Unruh-deWitt detectors located at $\mathbf{x}_1$ and $\mathbf{x}_2$, with a scattering interaction
\begin{align}
    H_{i,det} = -\lambda \int dt d^3\mathbf{x} \chi(t)g(\mathbf{x}-\mathbf{x}_i):\Psi(x)^2: \hat{\mu}(t)
\end{align}
with $g(\mathbf{x})$ a spatial smearing function of compact support, $\chi (t)$ a time switching function and   
\begin{align}
    \hat{\mu}(t) = e^{i\delta_1 t}b^\dagger + e^{-i\delta_1 t}b,
\end{align}
where $\delta_1,\delta_2$ are the detector energies, assumed small. Take an incoming particle state
\begin{align}
    |f\rangle = \int \frac{d^3\mathbf{k'}}{\sqrt{(2\pi)^32E_k'}}f_{\mathbf{k}}(k')a^\dagger_{k'}|0\rangle
\end{align}
such that $\langle f_k | f_k\rangle = 1$ and $a^\dagger_{k'}$ is the creation operator for the $\Psi$-field. The outgoing particle state is similarly $|h\rangle$, which we define as
\begin{align}
    |h\rangle = \int \frac{d^3\mathbf{k'}}{\sqrt{(2\pi)^32E_k'}}f_{\mathbf{p}}(k')a^\dagger_{k'}|0\rangle
\end{align}
i.e. it has the same structure but different central momentum $\mathbf{p}$.

We can then compute the Mott problem: what is the likelihood that a particle passes by the two detectors, exciting them, and continues on its way? In other words, we look for
\begin{align}
    \mathcal{A}&=\langle \mathrm{out}|\mathrm{in}\rangle,\\
    |\mathrm{in}\rangle &= |f\rangle\otimes |0\rangle _{\mathbf{x}_1}\otimes 0\rangle _{\mathbf{x}_2},\\
    |\mathrm{out}\rangle &= |h\rangle \otimes |\Delta_1 \rangle _{\mathbf{x}_1} \otimes |\Delta_2 \rangle _{\mathbf{x}_1}.
\end{align}
Calculating to second order in the interaction picture with $U = T\exp (-i\sum _i\int _{-\infty}^\infty H_{i,det}(t)dt)$ we obtain
\begin{align}
    \mathcal{A} ^{(2)} &= -\lambda^2 \int dt dt' \chi (t) \chi (t')e^{i\Delta_1 t + i\Delta _2 t'}\mathcal{M}(t,\mathbf{x}_1;t',\mathbf{x}_2) + (1 \leftrightarrow 2)\\
    \mathcal{M}(t,\mathbf{x}_1;t',\mathbf{x}_2)&:= \int d^3\mathbf{x}d^3\mathbf{x}' g(\mathbf{x}-\mathbf{x}_1)g(\mathbf{x}-\mathbf{x}_2)\langle h| T:\Psi(t,\mathbf{x})^2::\Psi (t',\mathbf{x}')^2:|f\rangle. \label{eq:wick}
\end{align}
We also obtain
\begin{align}
    &\langle h| T:\Psi(t,\mathbf{x})^2::\Psi (t',\mathbf{x'})^2:|f\rangle \nonumber\\
    &= 4\Delta_F (x-x')(H^*(x)F(x')+H^*(x')F(x)) + 2\Delta_F(x-x')^2\langle h|f\rangle \label{eq:wickopen}
\end{align}
with $\Delta _F$ the Feynman propagator and
\begin{align}
    F(x') = \langle 0|\Psi (x')|f\rangle, \quad H(x) = \langle 0 | \Psi (x)|h\rangle .
\end{align}
The second term in \eqref{eq:wickopen} is a vacuum term; the first one exchanges momentum between the detectors. Let us therefore focus on the first term.

The exchange amplitude is
\begin{align}
    &\mathcal{A}^{\text{exch}} \nonumber \\
    &= -\lambda^2 \int dt dt' \chi (t) \chi (t')e^{i\Delta_1 t + i\Delta _2 t'}\int d^3\mathbf{x}d^3\mathbf{x}'g(\mathbf{x}-\mathbf{x}_1)g(\mathbf{x}-\mathbf{x}_2)H^*(x)F(x')\Delta_F(x-x') .
\end{align}
Now, supposing that $\mathbf{R} \gg a$ with $a$ the size of the support of $g$. Then, if $g$ are strongly localized,
\begin{align}
    \Delta _F(x-x') \approx \Delta _F (t-t',\mathbf{x}_1-\mathbf{x}_2)
\end{align}
Then take the wave packets to be Gaussians in the semi-classical regime. In that case,
\begin{align}
    F(t,x) &\approx  C_f
  e^{-iE_k t+i\mathbf k\cdot\mathbf x}
  e^{-\sigma_f^2|\mathbf x-\mathbf v_k t|^2}\\
  H(t,\mathbf x)
  &\approx C_h
  e^{-iE_p t+i\mathbf p\cdot\mathbf x}
  e^{-\sigma_h^2|\mathbf x-\mathbf v_p t|^2}
\end{align}
with the wave packet widths $\sigma_h, \sigma_f$, velocities $\mathbf{v}_{q}=\mathbf{q}/E_{q}$ and $C_i$ the normalization constants, which may depend on time. Then
\begin{align}
    \bigg|\int d^3\mathbf{x}d^3\mathbf{x}'g(\mathbf{x}-\mathbf{x}_1)g(\mathbf{x}-\mathbf{x}_2)H^*(x)F(x')\Delta_F(x-x')\bigg|\nonumber\\
    \approx |\Delta_F(t-t',\mathbf x_1-\mathbf x_2)|
e^{-\sigma_f^2|\mathbf x_1-\mathbf v_k t|^2
   -\sigma_h^2|\mathbf x_2-\mathbf v_p t'|^2}
\end{align}
This is only appreciable when $(t',\mathbf{x}_2)-(t,\mathbf{x}_1)$ is not spacelike and $\mathbf{x}_1 \approx v_\mathbf{k}t$, $\mathbf{x}_2\approx v_\mathbf{p}t'$. Interpreting the situation as one particle passing through both detectors, we can assume that $\mathbf{v}_p \approx \mathbf{v}_k$ if the detector energies are not high. Then we obtain that the integral is large only when
\begin{align}
    \mathbf{x}_1-\mathbf{x}_2 \approx \mathbf{v}_k\Delta t
\end{align}
which approximately fixes the direction of $\mathbf{v}_k$ and the magnitude of $\Delta t$. 

We can use this result to improve momentum reconstruction. This would provide an exponential filter to the calculation in \ref{sec:3d}, where those values of $\psi$ which do not correspond to vectors near-parallel to $\mathbf{r}$ would be suppressed. Thus, by treating the detector system as semi-classical and applying a filtering process, we can even further constrain the allowed momenta. This argument is essentially similar to that applied by Mott to the bubble chamber problem \cite{mottWaveMechanicsRay1929}.
\section{Discussion}\label{sec:discussion}
We have investigated the applicability of Unruh-deWitt detector systems for momentum reconstruction in a scattering process with two out-state particles. In the case of two dimensions, a special case of the preceding calculation, we could determine all the momentum components fully; in the case of three dimensions, five of the six unknown momentum components could be reconstructed, and for the remaining component a conditional distribution was derived. 

Some notes about our assumptions are in order. First of all, we conditioned our calculation on both detectors becoming excited, and we further assumed that the only particles present were from the particle process under study. Finally, we assumed that the particle process and the detection process happen independently, even though both processes were taken to be asymptotic.

The first assumption simply means that we discard any event in which we did not have a double hit to the detectors. This seems reasonable, as the output of our particle process consists of two particles. It is also not problematic to assume that only particles from the decay process are present, since one could perform the experiment in a vacuum chamber. More problematic is the assumption of asymptotic time in both the detector and the particle process. Strictly speaking, if the processes overlap in time, we should take in to account the backreaction of the detector on the fields, which would then influence the decay amplitudes. This assumption thus amounts to saying that the distance between the detector and the decay event is large as compared with the energy scale of the decay event itself.

We also chose an unusual coupling for the Unruh-deWitt detectors and fields. Let us suppose we chose the conventional coupling instead, i.e.
\begin{align}
    H_{\text{det},i} &= -\epsilon_i \int dt \mu_i (t) \Psi (x_i,t),\\
    \mu _i(t) &=  e^{i\Delta _it}(b^{(i)})^\dagger + e^{-i\Delta _i t}b^{(i)}.
\end{align}
With this  choice, we are lacking a cross-term with both detector locations; the cross-terms containing $\delta$-functions with differing energy supports are zero unless $\Delta_1 = \Delta _2$. Thus \textit{there would be no dependence on the locations of the detectors whatsoever in the probability}. Consequently, the distribution for $p(\psi | \text{both excited},p)$ would be uniform in $\psi$, i.e. we would have no knowledge about $\psi$ at all. In our setup, information erasure actually improves our ability to reconstruct the momenta.

This can be intuitively understood in the following way. If the conventional coupling was used, then for each momentum $k_i$ we would obtain perfect knowledge of the particle's location by using the point-like detector, and thus by Heisenberg's uncertainty principle, we have no knowledge of its momentum. If we used an extended detector, the distribution would be the product of normalized form functions $\mathcal{F}_i$, and thus some momentum information could plausibly be obtained. Conversely, erasing the which-way information removes our ability to distinguish which particle hit which detector, and thus restores some of our ability to distinguish momenta.

Let us remark on the case of more than 2 outgoing particles. For $N\geq 2$ detectors (and $N$ outgoing particles) in $D$ spatial dimensions, we always have $N + D$ conservation laws. We need to work out $DN$ momentum components, i.e. full reconstruction is only possible in the case $DN \leq N+D$, which happens only for $N= 2$ and $D\leq 2$. For a higher number of particles, it would be possible to work out a conditional probability distribution, but increasing the number of particles leads to an increasing number of underdetermined components. Thus, the Unruh-deWitt detector setup is not well-suited to investigating outgoing states of more than two particles. It is also noteworthy that just as Unruh-deWitt detectors can be used to absorb particles, they can also be used to generate particle states; we briefly considered this in \cite{huhtalaNormalizingFockSpace2026}.

Finally, if for some reason the particle process does not e.g. conserve momentum or energy, then we would be left with more undetermined momentum components. This would be the case in, for example, curved spacetimes.

In conclusion, we have investigated using Unruh-deWitt detectors for momentum reconstruction in the context of particle processes and found that they can do a surprisingly well: using just the bare detector setup, we can obtain a distribution that far outperforms the uniform distribution. The properties of the investigated setup are non-trivial, and there exists an optimal setup for given sets of parameters. In further work, it would be interesting to investigate more elaborate detector setups and whether they could improve the obtained distribution even further.

\bibliography{apssamp}
\appendix
\section{Technical details}\label{app:technical}
\subsection{Unruh-deWitt interference}\label{app:interference}
In this subsection, we will calculate the transition amplitude
\begin{align}
    \mathcal{A}(k_1k_2\mapsto \Pi) = \langle \Pi,\text{out} |U_{\text{det}}|k_1k_2,\text{in}\rangle 
\end{align}
where 
\begin{align}
    |\Pi\rangle = \frac{1}{\sqrt{2}}\bigg[ |\Delta_1\rangle _{x_1}|\Delta_2\rangle _{x_2} + |\Delta_2\rangle_{x_1}|\Delta_1\rangle_{x_2} \bigg]
\end{align}
and
\begin{align}
    U_{\text{det}} &= e^{-i\lambda \sum _iH_{det,i}},\\
    H_{\text{det},i} &= \lambda \sum _{i=1,2}\mu _i(t)\chi_i(t)\Psi(x_i,t).
\end{align}
We take $\chi_i(t)$ to be switching functions, for example, a Gaussian wave packet. We have also chosen $\epsilon_1=\epsilon_2=\lambda$ with no loss of generality. We will take the squares of these switching functions as Dirac delta functions at the end.

Then to lowest order
\begin{align}
    \mathcal{A}(k_1k_2\mapsto \Pi) &\approx (-i)^2\lambda^2\int_{-\infty}^\infty dt \int_{-\infty}^\infty dt' \chi_1(t)\chi_2(t')\bigg[ \langle \Pi| T[H_{\text{det},1}(t)H_{\text{det},2}(t')]  |k_1k_2\rangle  \bigg]\\
    &=(-i)^2\lambda^2\int_{-\infty}^\infty dt \int_{-\infty}^\infty dt'\chi_1(t)\chi_2(t') e^{-i(E_{k_1}-\Delta_1)t -i(E_{k_2}-\Delta_2)t'+i\mathbf{k}_2 \mathbf{x}_2+i\mathbf{k}_1\cdot\mathbf{x}_1}\\
    &+(-i)^2\lambda^2\int_{-\infty}^\infty dt \int_{-\infty}^\infty dt'\chi_1(t)\chi_2(t') e^{-i(E_{k_1}-\Delta_2)t -i(E_{k_2}-\Delta_1)t'+i\mathbf{k}_2 \mathbf{x}_1+i\mathbf{k}_1\cdot\mathbf{x}_2}.
\end{align}
Write $\Phi _{12} = \mathbf{k}_1\cdot \mathbf{x}_1 + \mathbf{k}_2\cdot \mathbf{x}_2$ and $\Phi _{21} = \mathbf{k}_1\cdot \mathbf{x}_2 + \mathbf{k}_2\cdot \mathbf{x}_1$. We also write $E_{k_i} = E_i$ and 
\begin{align}
    \tilde{\chi}(\Omega ) = \int _{-\infty}^\infty dt \chi (t)e^{i\Omega t}.
\end{align}
Then
\begin{align}
    &\mathcal{A}(k_1k_2 \mapsto \Pi ) \nonumber\\
    &\approx - \frac{\lambda^2}{\sqrt{2}} \bigg( \tilde{\chi}(E_1-\Delta_1)\tilde{\chi}(E_2-\Delta_2) + \tilde{\chi}(E_1-\Delta_2)\tilde{\chi}(E_2-\Delta_1)\bigg)(e^{i\Phi_{12}}+ e^{i\Phi_{21}}).
\end{align}
Squaring and taking the $\chi_i$ to the delta function limit, we obtain
\begin{align}
    |\mathcal{A}(k_1k_2\mapsto \Pi)|^2 = \lambda^4 &\bigg[ \delta(E_1-\Delta_1)\delta(E_2-\Delta_2) + \delta(E_1-\Delta_2)\delta(E_2-\Delta_1) \bigg]\\
    \times&(2+2\cos ((\mathbf{k}_1-\mathbf{k}_2)\cdot (\mathbf{x}_1-\mathbf{x}_2)).
\end{align}

\subsection{$\psi$-distribution}\label{app:distribution}
In this section, we will calculate the final form of the $\psi$ distribution. We start from the form of the problem after momentum conservation has been used. The integral to be computed is then
\begin{align}
I := \frac{\delta(\Delta E)}{(2\pi)^3\omega_p} \int \frac{1}{(2\pi)^32\sqrt{m^2+(\mathbf{p}-\mathbf{k}_1)^2}}\dot{p}(\Pi | k_1(p-k_1))(2\pi)^4|\mathcal{M}|^2 \widetilde{d^3k}
\end{align}
where $\delta(\Delta E)$ is the total energy conservation in the particle process, i.e. $\Delta E = \omega _p - \Delta_1 - \Delta_2$. We now make the assumption that $\mathcal{M}$ does not depend on $\psi$ to obtain the simple analytical form \eqref{eq:distributionexplicit}. Then
\begin{align}
    I = \frac{\delta (\Delta E)|\mathcal{M}|^2\mathcal{N}}{(2\pi)^5\omega _p}\sum_{(i,j)=(1,2),(2,1)}\int \frac{d^3k_1}{4E_{k_1}E_{\mathbf{p}-\mathbf{k}_1}}K(\mathbf{x}_1,\mathbf{x}_2,\mathbf{k}_1,\mathbf{p}-\mathbf{k}_1)\delta (E_{k_1}-\Delta_i)\delta(E_{k_2}-\Delta_j).
\end{align}
To compute this integral, we will now switch to spherical coordinates. Let us first choose our coordinate axes such that $\mathbf{p}$ is parallel to the z-axis, $\mathbf{p}\parallel \hat{\mathbf{z}}$. We will write $\mathbf{r}=\mathbf{x}_1-\mathbf{x}_2$, $r=|\mathbf{r}|$, $P = |\mathbf{p}|$ and $k = |\mathbf{k}_1|$. Then, in spherical coordinates, $\theta$ is the (polar) angle between $\mathbf{k}_1$ and $\mathbf{p}$, and $\psi\in [0,2\pi]$ is the azimuthal angle. Thus
\begin{align}
    d^3k_1 &= k^2dk d(\cos \theta)d\psi ,\\
    E_{\mathbf{p}-\mathbf{k}_1} &= \sqrt{m^2+P ^2+k^2-2Pk\cos \theta }.
\end{align}
Hence
\begin{align}
    I = \frac{\delta (\Delta E)|\mathcal{M}|^2\mathcal{N}}{(2\pi)^5\omega _p}\sum_{(i,j)}\int _0^\infty dk \int_{-1}^1d\cos\theta \int_0^{2\pi}&\frac{k^2}{4E_{k}E_{\mathbf{p}-\mathbf{k}_1}}K\delta(E_k-\Delta_i)\delta(E_{\mathbf{p}-\mathbf{k}_1}-\Delta_j) .
\end{align}
where we have omitted the argument in $K$. Now we can use the delta functions. Let us write
\begin{align}
    \kappa_i = \sqrt{\Delta _i^2 - m^2}, \quad i=1,2.
\end{align}
We have
\begin{align}
    E_k = \sqrt{m^2+k^2}\implies \frac{dE_k}{dk} = \frac{k}{E_k}.
\end{align}
Thus $\delta (E_k - \Delta _i) = \frac{\Delta _i}{k}\delta (k-\kappa _i)$ and we get from the $k$-integral
\begin{align}
    I = \frac{\delta (\Delta E)|\mathcal{M}|^2\mathcal{N}}{(2\pi)^5\omega _p}\sum_{(i,j)=(1,2),(2,1)} \int _{-1}^1d\cos \theta \int_0^{2\pi} \frac{\kappa_i}{4E_{\mathbf{p}-\mathbf{k}_1}}K\bigg|_{k=\kappa _i} \delta (E_{\mathbf{p}-\mathbf{k}_1}-\Delta _j).
\end{align}
Note that we must assume $\kappa_i > m$. Now
\begin{align}
    E_{\mathbf{p}-\mathbf{k}_1} = \sqrt{m^2 + P^2 + \kappa_i^2 - 2P\kappa _i\cos (\theta)},
\end{align}
and thus from the delta-function:
\begin{align}
    \delta (E_{\mathbf{p}-\mathbf{k}_1} - \Delta _j) = \frac{\delta (\cos \theta - \cos\theta_{ij})}{|dE_{\mathbf{p}-\mathbf{k}_1}/d\cos\theta|_{\theta = \theta _{ij}}} = \frac{E_{\mathbf{p}-\mathbf{k}_1}}{P\kappa _i} \delta (\cos \theta - \cos \theta_{ij}).
\end{align}
Here, we can find $\cos \theta _{ij}$ using $\Delta _j^2 = \kappa_j^2+m^2$, we obtain
\begin{align}
    \cos \theta _{ij} = \frac{P^2 + \kappa_i^2 - \kappa_j^2}{2P\kappa_i}.
\end{align}
Here, we must of course assume $|\cos \theta _{ij}|\leq 1$. Then we finally have
\begin{align}
    I=\frac{\delta (\Delta E)|\mathcal{M}|^2\mathcal{N}}{(2\pi)^5\omega _p4P}\sum_{(i,j)=(1,2),(2,1)}\Theta (\Delta_i - m)\Theta (1-|\cos \theta_{ij}|)\int _0^{2\pi}K_{ij}(\psi).
\end{align}
Let us now be explicit about the form of $K$. Recall that
\begin{align}
    K(\mathbf{x}_1,\mathbf{x}_2,\mathbf{k}_1,\mathbf{p}-\mathbf{k}_1) = 2 + 2\cos ([2\mathbf{k}_1-\mathbf{p}]\cdot \mathbf{r})
\end{align}
Let us denote the angle between $\mathbf{p}$ (which is parallel to the z-axis) and $\mathbf{r}$ as $\alpha$. Then after a rotation of coordinate axes we can write
\begin{align}
    \mathbf{r} = r (\sin \alpha, 0, \cos \alpha)
\end{align}
and
\begin{align}
    \mathbf{k}_1 = k(\sin \theta_{ij} \cos \psi, \sin \theta_{ij}\sin \psi, \cos \theta_{ij}).
\end{align}
Then 
\begin{align}
    (2\mathbf{k}_1-\mathbf{p})\cdot \mathbf{r} = 2kr(\cos \theta_{ij}\cos \alpha + \sin\theta_{ij}\sin\alpha \cos \psi) - Pr \cos \alpha.
\end{align}
Thus
\begin{align}
    K_{ij}(\psi) &= 2 + 2\cos (A_{ij}+B_{ij}\cos \psi), \\
    A_{ij} &= 2\kappa _i r \cos \theta_{ij} \cos \alpha - Pr \cos \alpha, \\
    B_{ij} &= 2\kappa _i r \sin \theta_{ij}\sin \alpha
\end{align}
and hence
\begin{align}
    I = \frac{\delta (\Delta E)|\mathcal{M}|^2\mathcal{N}}{(2\pi)^5\omega _p4P}\sum_{(i,j)=(1,2),(2,1)}&\Theta (\Delta_i - m)\Theta (1-|\cos \theta_{ij}|)\\
    \times &\int _0^{2\pi}\bigg[2 + 2\cos (A_{ij}+B_{ij}\cos \psi)\bigg].
\end{align}
The integrand, along with the associated Heaviside functions, is the distribution, and the full integral is the normalization; all the prefactors cancel upon normalization.
\end{document}